\documentclass[10pt,conference]{IEEEtran}
\usepackage{graphicx}
\usepackage{amsmath}
\usepackage{amssymb}
\usepackage{amsfonts}

\newcommand{\twobibs}[2]{#2} 

\newcommand{\myincludegraphics}[2]{\includegraphics{#1.#2}}

\newcommand{\thesisonly}[1]{}
\newcommand{\paperonly}[1]{#1}
\newcommand{\bbcaption}[3]{\caption{#2}}

\newcommand{\sq}{\mathrm{sq}}
\newcommand{\bfzero}{\mathbf{0}}

\newcommand{\Prob}{\mathrm{Prob}}

\newcommand{\Bis}{\mathsf{B}} 
\newcommand{\Dis}{\mathsf{B}} 
\newcommand{\Iis}{\mathsf{U}} 
\newcommand{\Uis}{\mathsf{U}} 
\newcommand{\Tis}{\mathsf{T}}

\newcommand{\integers}{\mathbb{Z}}
\newcommand{\size}[1]{\left|#1\right|}
\newcommand{\set}[1]{\left\{#1\right\}}
\newcommand{\bfpi}{\boldsymbol{\pi}}

\newcommand{\bfdelta}{\boldsymbol{\delta}}

\newcommand{\wasNabla}{\Dis}
\newcommand{\wasBarNabla}{\Iis}

\newcommand{\calS}{\mathbb{S}}
\newcommand{\capacity}{\mathsf{cap}}
\newcommand{\encoder}{\mathcal{E}}
\newcommand{\lpmax}{\mathrm{lp}_\mathrm{max}}
\newcommand{\lpmin}{\mathrm{lp}_\mathrm{min}}

\newcommand{\Footnotemark}[1]{${}^{#1}$}
\newcommand{\Footnotetext}[2]{\begin{figure}[b]\footnotesize%
  \vspace{-3ex}\hrulefill\hfill\makebox[0em]{}\hfill\makebox[0em]{}%
  \vfill
  \par${}^{#1}$ #2\vspace{-0.60ex}\end{figure}\addtocounter{figure}{0}}

\newtheorem{theo}{Theorem}
\newtheorem{lemm}[theo]{Lemma}
\newtheorem{coro}[theo]{Corollary}

\begin{document}

\title{Bounds on the Rate of 2-D Bit-Stuffing Encoders\Footnotemark{*}}

\author{\IEEEauthorblockN{Ido Tal \qquad Ron M. Roth}
\IEEEauthorblockA{
Computer Science Department\\
Technion, Haifa 32000, Israel.\\
Email: \{{\tt idotal, ronny}\}{\tt @cs.technion.ac.il}
}}
\maketitle

\begin{abstract}
A method for bounding the rate of bit-stuffing encoders for 2-D constraints is presented. Instead of considering the original encoder, we consider a related one which is quasi-stationary. We use the quasi-stationary property in order to formulate linear requirements that must hold on the probabilities of the constrained arrays that are generated by the encoder. These requirements are used as part of a linear program. The minimum and maximum of the linear program bound the rate of the encoder from below and from above, respectively.

A lower bound on the rate of an encoder is also a lower bound on the capacity of the corresponding constraint. For some constraints, our results lead to tighter lower bounds than what was previously known.

\end{abstract}
\Footnotetext{*}{This work was supported by grant
       No.~2002197 from
       the United-States--Israel Binational Science Foundation (BSF),
       Jerusalem, Israel.
The results of this work were presented at the IEEE International Symposium on Information Theory, Toronto, Ontario, Canada, July 2008.
}
%
%
\thesisonly{\bchapter{חסמים על קצבים של מקודדי \L{bit-stuffing} דו-מימדיים}{Bounds on the Rate of 2-D Bit-Stuffing Encoders}
\label{chap:linear}}%
\paperonly{\section{Introduction}}%
\thesisonly{\bsection{מבוא}{Introduction}}%
Two-dimensional (2-D) constraints are formally defined in \cite{HalevyRoth:02}. Consider a 2-D constraint $\calS$ defined over some finite alphabet $\Sigma$. Informally, a bit-stuffing encoder for $\calS$ operates as follows. We encode information to an $M \times N$ rectangular array; namely, we produce an array $a \in \calS \cap \Sigma^{M \times N}$. We first initialize the ``boundaries'' of the array (formally defined later) according to some fixed probability distribution. Then, we write to the ``interior'' of the array in raster fashion: row-by-row. The symbol currently written is the result of a coin toss. The probability distribution of the coin is a function of neighboring symbols, which have already been written. However, the ``coins'' used are in fact (invertible) probability transformers, the input of which is the information we wish to encode. Thus, information can be encoded, and decoded.

A bit-stuffing encoder is ``variable-rate''. The bit-stuffing technique was initially devised for encoding one-dimensional (1-D) constraints \cite{BenderWolf:93}. In \cite{Halevy+:04} and \cite{RothSiegelWolf:01}, bit-stuffing encoders for specific 2-D constraints were presented and analyzed. In \cite{ForchhammerLaursen:07}, a slightly different definition of bit-stuffing was used to give lower bounds on the capacity of specific 2-D constraints.

In this work, we derive upper and lower bounds on the rate of a general bit-stuffing encoder. A lower bound on the rate of an encoder is also a lower bound on the capacity of the corresponding constraint:
\[
\capacity(\calS) = \lim_{M,N \to \infty}
\frac{1}{M\cdot N} \cdot \log_2 \size{\calS \cap \Sigma^{M \times N}} \; .
\]
For some constraints, our results lead to tighter lower bounds on capacity than what was previously known.

Fix some 2-D constraint $\calS$ over an alphabet $\Sigma$. As a running example, consider the kings constraint $\calS_\sq$, defined over the binary alphabet $\Sigma_\sq=\set{0,1}$ (see Figure~\ref{fig:example:square}). A binary array satisfies the kings constraint if each entry set to ``1'' has all of its eight-neighbors set to ``0''. Namely, two entries equal to ``1'' may not appear consecutively along a row, column, or diagonal.

\begin{figure}
\[
\newcommand{\zero}{{\textrm{\scriptsize $0$}}}
\newcommand{\one}{{\textrm{\scriptsize $1$}}}
\renewcommand{\bfzero}{{\textrm{\scriptsize $\mathbf{0}$}}}
\renewcommand{\arraystretch}{0.7}
 \arraycolsep0.5ex
\begin{array}{cccccccc}
\one & \bfzero & \zero & \zero & \bfzero & \one & \bfzero & \zero\\
\bfzero & \bfzero & \zero & \zero & \bfzero & \bfzero & \bfzero & \bfzero \\
\zero & \zero & \zero & \zero & \zero & \bfzero & \one & \bfzero \\
\bfzero & \bfzero& \bfzero & \zero & \zero & \bfzero & \bfzero & \bfzero \\
\bfzero & \one & \bfzero & \zero & \zero & \zero & \zero & \zero
\end{array}
 \]
\bbcaption{מערך בינארי המקיים את אילוץ המלכים.}{Binary array satisfying the kings constraint. If we flip any one
(or more) of the highlighted ``0'' bits to ``1'', then the resulting array will
not satisfy the kings constraint.}{Binary array satisfying the kings constraint.}
\label{fig:example:square}
\end{figure}

The rest of this paper is organized as follows. In Sections \ref{sec:notation} and \ref{sec:bitStuffer}, we define our notation and our model of a bit-stuffing encoder, respectively. In Section \ref{sec:quasiStationarity}, we define the concept of quasi-stationarity. We also prove that, w.l.o.g., we may assume that our encoder is quasi-stationary. In Section \ref{sec:linearProgram}, we take advantage of the quasi-stationary property and define a linear program. The minimum (maximum) of the linear program bounds the rate of our encoder from below (above). Finally, section \ref{sec:lowerBoundOnCapacity} states a generic lower bound on capacity, and contains examples where this bound improves on previous results.

We note at this point that although this
work deals with 2\hbox{-}D constraints, our method can be easily generalized to higher dimensions as well.

\paperonly{\section{Notation}}%
\thesisonly{\bsection{סימונים}{Notation}}%
\label{sec:notation}
We first set up some notation.

\begin{trivlist}
\item[\textbf{Parallelogram and rectangle:}] For $M, N > 0$ and $t \geq 0$, denote
\[
\Bis^{(t)}_{m,n} = \set{(i,j) : 0 \leq i < M \; , \quad 0 \leq t \cdot i + j < N} \; .
\]
Also, for $t = 0$, denote
\[
\Bis_{M,N} = \Bis^{(0)}_{m,n} \; .
\]
\item[\textbf{Configuration:}]Let $a = (a_{i,j})_{(i,j) \in \Uis}$ be a 2-D configuration over $\Sigma$. Namely, the index set satisfies $\Uis \subseteq \integers^2$, and for all $(i,j) \in \Uis$ we have that $a_{i,j} \in \Sigma$.

\item[\textbf{Shifts:}]
For integers $\alpha, \beta$ we denote the shifted index set as
\[
\sigma_{\alpha,\beta}(\Uis) = \set{(i+\alpha,j+\beta) : (i,j) \in \Uis} \; .
\]
Also, by abuse of notation,
let $\sigma_{\alpha,\beta}(a)$ be the shifted configuration (with
index set $\sigma(\Uis)$):
\[
\sigma_{\alpha,\beta}(a)_{i+\alpha, j + \beta} = a_{i,j} \; .
\]

\item[\textbf{Restriction of configuration:}]
For an index set $\Psi \subseteq \Uis$, denote the restriction of $a$ to $\Psi$ by $a[\Psi] = (a[\Psi]_{i,j})_{(i,j) \in \Psi}$. Namely,
\[
a[\Psi]_{i,j} = a_{i,j} \; , \quad \mbox{where} \quad (i,j) \in \Psi \; .
\]

\item[\textbf{Shift and restrict:}]
Let $\tau_{\alpha,\beta}(a,\Psi)$ be shorthand for
\[
\tau_{\alpha,\beta}(a,\Psi) = (\sigma_{-\alpha,-\beta}(a))[\Psi] \; .
\]
Namely, shift the configuration $a$ such that index $(\alpha,\beta)$ is now index $(0,0)$, and then restrict to $\Psi$.

\item[\textbf{Boundary:}] Denote by $\partial(\Uis,\Psi)$ the set of all the indexes $(\alpha,\beta) \in \Uis$ for which the ``shift and restrict'' operation is invalid.
\[
\partial(\Uis,\Psi) = \set{(\alpha,\beta) \in \Uis : \sigma_{\alpha,\beta}(\Psi) \not\subseteq \Uis } \; .
\]
The index set $\partial(\Uis,\Psi)$ is termed the ``boundary'', and the ``interior'' is
\[
\bar\partial(\Uis,\Psi) = \Uis \setminus \partial(\Uis,\Psi) \; .
\]
When $\Uis=\Bis_{M,N}$ and $\Psi$ is understood from the context, we abbreviate
\[
\partial_{M,N} = \partial(\Bis_{M,N},\Psi) \; , \quad \bar\partial_{M,N} = \bar\partial(\Bis_{M,N},\Psi) \; .
\]
 Figure~\ref{fig:boundary} shows an example of such sets, where
\begin{equation}
\label{eq:PsiRunning}
\Psi=\set{(0,-2),(0,-1),(-1,-1),(-1,0),(-1,1)} \; .
\end{equation}

\item[\textbf{Restriction of constraint:}]
Denote the restriction of $\calS$ to $\Uis$ by
\[
\calS[\Uis] = \set{ a : \mbox{there exists $a' \in \calS$ such that $a'[\Uis] = a$}} \; .
\]
If $\Uis=\Bis_{M,N}$, then we abbreviate
\[
\calS_{M,N} = \calS[\Bis_{M,N}] \; .
\]

\item[\textbf{Lexicographic ordering:}]
We define a lexicographic ordering $\prec$ on $\integers^2$ as
\[
(i',j') \prec (i,j) \quad \Longleftrightarrow \quad (i' < i) \;\;
\mbox{or} \;\; (i'=i \quad \mbox{and} \quad j' < j) \; .
\]
Also, we define the index set
\begin{equation}
\label{eq:Tij}
\Tis_{i,j} = \set{(i',j'):(i',j') \prec (i,j)} \; .
\end{equation}
\end{trivlist}

\begin{figure}[ht]
\centering
\myincludegraphics{linear}{1}
\bbcaption{קבוצות האינדקסים $\Psi$, $\Bis_{M,N}$, $\partial_{M,N}$, ו- $\bar\partial_{M,N}$.}{The index $(0,0)$ is represented by $\bullet$.
We take $\Psi$ as in (\ref{eq:PsiRunning}), and it is represented by the diagonally striped cells. We set $M = 5$ and $N = 8$. The index set $\Bis_{M,N}$ is represented by the shaded part (both light and dark). The boundary $\partial_{M,N}$ is represented by the lighter shaded part, while the interior $\bar\partial_{M,N}$ is represented by the darker shaded part.}{Index sets $\Psi$, $\Bis_{M,N}$, $\partial_{M,N}$, and $\bar\partial_{M,N}$.}
\label{fig:boundary}
\end{figure}

\paperonly{\section{Bit stuffer definitions}}%
\thesisonly{\bsection{הגדרות של מקודדי \L{bit-stuffing}}{Bit stuffer definitions}}%
\label{sec:bitStuffer}
In this section, we present the formal definition of bit-stuffing encoders. A bit-stuffing encoder for $\calS$ is defined through 
a
triple
\[
\encoder = (\Psi,\mu,\bfdelta=(\delta_{M,N})_{M,N>0}) \; .
\]
The set
\begin{equation}
\label{eq:neighboursSet}
\Psi \subseteq \Tis_{0,0}
\end{equation}
is termed the \emph{neighbor set}. The \emph{conditional probability function} $\mu$,
\[
\mu( \cdot | \cdot) \; , \quad \mu : \Sigma \times \calS[\Psi] \to [0,1] \; ,
\]
is a conditional probability distribution on $\Sigma$, given an element of $\calS[\Psi]$. For $M,N>0$, the \emph{boundary probability function}
\[
\delta_{M,N} : \calS[\partial_{M,N}] \to [0,1]
\]
is a probability distribution on $\calS[\partial_{M,N}]$. From here onward, we fix $\encoder$.

For our running example, let the neighbor set $\Psi_\sq=\Psi$ be as in (\ref{eq:PsiRunning}),
and define $\varphi^{(0)},\varphi^{(1)} \in \calS_\sq[\Psi]$ as
\[
\begin{array}{ccccc}
\varphi^{(0)}_{0,-2} {=} 0 & \varphi^{(0)}_{0,-1} {=} 0 & \varphi^{(0)}_{-1,-1} {=} 0 & \varphi^{(0)}_{-1,0} {=} 0 & \varphi^{(0)}_{-1,1} {=} 0 \\
\varphi^{(1)}_{0,-2} {=} 1 & \varphi^{(1)}_{0,-1} {=} 0 & \varphi^{(1)}_{-1,-1} {=} 0 & \varphi^{(1)}_{-1,0} {=} 0 & \varphi^{(1)}_{-1,1} {=} 0
\end{array}
\]
(see Figure~\ref{fig:runningNeighborSet}). Also, take the conditional probability function as
\begin{equation}
\label{eq:musq}
\mu_\sq(1|\varphi) = 1 - \mu_\sq(0|\varphi) =
\begin{cases}
0.258132 & \varphi = \varphi^{(0)}\\
0.312231 & \varphi = \varphi^{(1)}\\
0 & \mbox{otherwise} \; .
\end{cases}
\end{equation}
Thus, $\mu_\sq(\cdot|\cdot)$ can be implemented using two coins (one for the context $\varphi^{(0)}$ and one for $\varphi^{(1)}$). For our running example, we take $\delta_{M,N}$ as the function equal to $1$ for the all zero boundary $(0)_{(i,j) \in \partial_{M,N}}$, and $0$ for all other members of $\calS_\sq[\partial_{M,N}]$.
\begin{figure}
\[
\varphi^{(0)} =
\begin{array}{cccc}
& 0 & 0 & 0\\
0 & 0 & \bullet
\end{array}
\qquad
\varphi^{(1)} =
\begin{array}{cccc}
& 0 & 0 & 0\\
1 & 0 & \bullet
\end{array}
\]
\bbcaption{שתי הקונפיגורציות הלא טריבאליות עבור $\mu$ בדוגמה הרצה}{The two non-trivial configurations for $\mu$ in our running example, where $\bullet$ designates coordinate $(0,0)$.}{The two non-trivial configurations for $\mu$ in the running example.}
\label{fig:runningNeighborSet}
\end{figure}

Given integers $M,N > 0$, the bit-stuffing encoder $\encoder$ defines a probability measure on the elements  $a = (a_{i,j})_{(i,j) \in \Bis_{M,N}}$ of $\Bis_{M,N}$, in the following manner.
As a first step, we set the boundary $a[\partial_{M,N}]$, according to the probability distribution $\delta_{M,N}$.
Next, we write the contents of the interior of $a$ in raster fashion: row-by-row, from left to right. The probability of writing $w \in \Sigma$ in entry $(i,j) \in \bar\partial_{M,N}$ is given by
\[
\Prob(a_{i,j} = w ) = \mu(w|(\tau_{i,j}(a, \Psi)) \; .
\]
Specifically, note that when writing entry $(i,j)$, we have by (\ref{eq:neighboursSet}) that $\tau_{i,j}(a)$ is a function of entries of $a$ which have already been written. A fundamental requirement for $\Psi$ and $\mu$ is that for every $M$, $N$, and $\delta_{M,N}$, the support of the probability measure thus defined is contained in $\calS_{M,N}$.

Let
\[
A(\encoder, M, N) = A = (A_{i,j})_{(i,j) \in \Bis_{M,N}}
\]
be a random variable taking values on $\calS_{M,N}$ according to the measure we have just defined. Namely,
\begin{multline}
\label{eq:Aprob}
\Prob(A=a) = \delta_{M,N}(a[\partial_{M,N}]) \cdot \\
\prod_{(i,j) \in \bar\partial_{M,N}} \mu(a_{i,j} | \tau_{i,j}(a,\Psi) ) \; .
\end{multline}

We now explain how $\encoder$ is used to actually encode information. The ``coin tosses'' corresponding to the invocations of $\mu$ are, in effect, a function of the information we wish to encode. Specifically, the values of the tosses are the output of distribution transformers on the input stream (the mapping from the input stream to the sequence of coin tosses is one-to-one) \cite{RothSiegelWolf:01}. Thus, we may encode information, and also decode it. So, we define the \emph{rate} of our encoder as
\[
R(\encoder) \triangleq \liminf_{M,N \to \infty} \frac{H(A[\bar\partial_{M,N}]|A[\partial_{M,N}])}{M \cdot N} \; ,
\]
where
\[
A = A(\encoder,M,N) \; .
\]

Note that since
\[
\liminf_{M,N \to \infty} \frac{\size{\bar\partial_{M,N}}}{M \cdot N} = 1 \; ,
\]
we also have that
\[
R(\encoder) = \liminf_{M,N \to \infty} \frac{H(A(\encoder,M,N))}{M \cdot N} \; .
\]
\vspace{5pt}
\paperonly{\section{Quasi-stationarity}}%
\thesisonly{\bsection{קואזי-סטציונריות}{Quasi-stationarity}}%
\label{sec:quasiStationarity}
Fix $k>0$.
Define the random variable
\[
A^{(k)}(\encoder,M,N) = A^{(k)}=(A^{(k)}_{i,j})_{(i,j) \in \Bis_{M,N}}
\]
taking values on $\calS_{M,N}$ as follows. For $w \in \calS_{M,N}$, we have
\[
\Prob(A^{(k)} {=} w) = \frac{1}{k^2}\sum_{0 \leq i,j < k} \Prob( \sigma_{-i,-j}(A'[\Bis_{M,N}]) {=} w) \; ,
\]
where
\[
A' = A(\encoder,M+k-1,N+k-1) \; .
\]
Namely, given $A'$, we randomly and uniformly pick an $M \times N$ sub-configuration of it, and shift accordingly. The usefulness of $A^{(k)}$ is that it is ``quasi-stationary'' \cite[\S 6]{Halevy+:04}.

\begin{lemm}[{\cite[Proposition 6.1]{Halevy+:04}}]
Let $\encoder$, $M$, $N$, and $k$ be given. Let $\Uis \subseteq \Bis_{M,N}$ be an index set, and let $w \in \calS[\Uis]$ be given. Suppose that for given integers $\alpha,\beta$ we have that $\sigma_{\alpha,\beta}(\Uis) \subseteq \Bis_{M,N}$. Denote $A^{(k)}=A^{(k)}(\encoder,M,N)$. Then,
\begin{multline*}
\size{\Prob (A^{(k)}[\Uis] = w) - \Prob(A^{(k)}[\sigma_{\alpha,\beta}(\Uis)] = \sigma_{\alpha,\beta}(w)) }
\\
\leq \frac{\size{\alpha}+\size{\beta}}{k} \; .
\end{multline*}
\end{lemm}

Next, we show that $A^{(k)}$ is a random variable corresponding to an encoder very similar to $\encoder$. First, define $\bfdelta^{(k)} = (\delta^{(k)}_{M,N})_{M,N>0}$, where
\[
\delta^{(k)}_{M,N} : \calS[\partial_{M,N}] \to [0,1]
\]
(that is, $\delta^{(k)}_{M,N}$ is a probability distribution on $\calS[\partial_{M,N}]$), and for every $d \in \calS[\partial_{M,N}]$,
\[
\delta^{(k)}_{M,N}(d) = \Prob(A^{(k)}(\encoder,M,N)[\partial_{M,N}]=d) \; .
\]
Next, define the encoder $\encoder^{(k)}$ as
\begin{equation}
\label{eq:Ek}
\encoder^{(k)} = (\Psi,\mu,\bfdelta^{(k)}) \; .
\end{equation}
\begin{lemm}[{\cite[Proposition 6.2]{Halevy+:04}}]
\label{lemm:AkversusAEk}
The probability distributions of $A^{(k)}(\encoder,M,N)$ and $A(\encoder^{(k)},M,N)$ are equal.
\end{lemm}

The next lemma essentially states that the normalized entropies of $A$ and $A^{(k)}$ are asymptotically equal (for $M,N \to \infty$ and $k$ fixed). The proof is straightforward.
\begin{lemm}
\label{lemm:REequalToREk}
Fix an integer $k>0$. Then,
\[
R(\encoder) = R(\encoder^{(k)}) \; .
\]
\end{lemm}

It follows from Lemma~\ref{lemm:REequalToREk} that we can obtain bounds on $R(\encoder)$ by bounding instead the rate of the quasi-stationary encoder $\encoder^{(k)}$. And, indeed, quasi-stationarity will turn out to be useful for this purpose.

\paperonly{\section{Linear program}}%
\thesisonly{\bsection{תוכנית לינארית}{Linear program}}%
\label{sec:linearProgram}
In this section, we present lower and upper bounds on $R(\encoder)$. The bounds will be expressed as values of corresponding linear programs.

For $r,s > 0$ and $ t \geq 0$, we say that the parallelogram $\Bis_{r,s}^{(t)}$ is valid with respect to the neighbor set $\Psi$ if the set
\begin{equation}
\label{eq:interiorForParallelogram}
\set{(\alpha,\beta) : \left( \Psi \cup (0,0) \right) \subseteq \sigma_{\alpha,\beta}(\Bis_{r,s}^{(t)}) }
\end{equation}
is non-empty. Namely, some shift of the parallelogram includes the neighbor set $\Psi$ and $(0,0)$. From here onward, we fix $r$, $s$, and $t$ so that $\Bis_{r,s}^{(t)}$ is valid. Also, we fix $u$ and $v$, where $(u,v)$ is the largest element of (\ref{eq:interiorForParallelogram}), with respect to the ordering $\prec$.

\begin{figure}[ht]
\centering
\myincludegraphics{linear}{2}
\bbcaption{קבוצות האינדקסים $\Psi$, $\Lambda$ ו- $\Gamma$.}{The index sets $\Psi$, $\Lambda$, and $\Gamma$. The index sets are shown for $r=4$, $s=5$, and for both $t=0$ and $t=1$. The index $(0,0)$ is represented by $\bullet$.
We take $\Psi$ as in (\ref{eq:PsiRunning}), and it is represented by the diagonally striped cells. The index set $\Lambda$ is represented by the shaded part (both light and dark). The boundary $\Gamma$ is represented by the lighter shaded part. Note that $\Psi \subseteq \Gamma \subseteq \Lambda$.
}{The index sets $\Psi$, $\Lambda$, and $\Gamma$.}
\label{fig:lambda}
\end{figure}

Denote (see Figure~\ref{fig:lambda})
\[
\Lambda = \sigma_{u,v}(\Bis_{r,s}^{(t)}) \; , \quad \Gamma = \partial(\Lambda,\Psi) \; .
\]
For an as yet unspecified probability distribution over $\calS[\Gamma]$
\[
\pi(z) \; , \quad z \in \calS[\Gamma] \; ,
\]
define the random variable $Y$ taking values on $\calS[\Lambda]$ as follows. For $y \in \calS[\Lambda]$,
\begin{equation}
\label{eq:ProbY}
\Prob( Y = y ) = \pi(y[\Gamma]) \prod_{(i,j) \in \Lambda \setminus \Gamma} \mu(y_{i,j} | \tau_{i,j}(y,\Psi) )
\end{equation}
(compare to (\ref{eq:Aprob})). Note that $\Prob( Y = y )$ is a linear function of the various $\pi(z)$'s. Next, define
\[
\Lambda' = \sigma_{u,v}(\Bis_{r-1,s}^{(t)}) \; , \quad \Lambda'' = \sigma_{u,v}(\Bis_{r,s-1}^{(t)}) \; ,
\]
and
\[
\Gamma' = \partial(\Lambda',\Psi) \; , \quad \Gamma'' = \partial(\Lambda'',\Psi) \; .
\]

Consider the linear program in Figure~\ref{fig:linearProgram}. First, note that it is indeed a linear program. Namely, recall that by (\ref{eq:ProbY}), the probability distribution of $Y$ is a linear function of the $\pi(z)$'s. Thus, both sides of (\ref{eq:hstationary}) and (\ref{eq:vstationary}) are also linear functions of the $\pi(z)$'s. For example, the LHS of (\ref{eq:hstationary}) equals
\[
\sum_{y \in \calS[\Lambda]:y[\Gamma']=z'} \pi(y[\Gamma]) \prod_{(i,j) \in \Lambda \setminus \Gamma} \mu(y_{i,j} | \tau_{i,j}(y,\Psi) ) \; .
\]
Denote the value of the linear program when minimizing by $\lpmin^{*}=\lpmin^{*}(\encoder)$, and when maximizing by $\lpmax^{*}=\lpmax^{*}(\encoder)$. Since (\ref{eq:Aprob}) and (\ref{eq:ProbY}) are very similar, we may intuitively say that $\encoder$ outputs $Y$. The optimization is over the probability distribution of the boundary $Y[\Gamma]$. The linear requirements (\ref{eq:hstationary}) and (\ref{eq:vstationary}) are added to force the distribution of $Y$ to be stationary. The objective function is the rate at point $(0,0)$.

The following theorem is our main result.
\begin{theo}
\label{theo:upperLowerBound}
For the linear program in  Figure~\ref{fig:linearProgram}, we have that
\[
\lpmin^{*} \leq R(\encoder) \leq \lpmax^{*} \; .
\]
\end{theo}
\paperonly{\begin{figure}[b]}
\thesisonly{\begin{figure}[t]}
\makebox[0ex]{}\hrulefill\makebox[0ex]{}
\vspace{1ex}
\small
\begin{multline*}
\!\!\!\!\!\!\!\textrm{Minimize (Maximize)} \\
-\sum_{z \in \calS[\Gamma]} \pi(z) \sum_{w \in \Sigma} \mu(w|z[\Psi]) \log_2 \mu(w|z[\Psi])
\end{multline*}
over the variables $(\pi(z): z \in \calS[\Gamma])$, subject to the following:
\[
\sum_{z \in \calS[\Gamma]} \pi(z) = 1 \; .
\]
For all $z \in \calS[\Gamma]$,
\[
\pi(z) \geq 0 \; .
\]
For all $z' \in \calS[\Gamma']$,
\begin{equation}
\label{eq:hstationary}
\Prob(Y[\Gamma'] = z') = \Prob(Y[\sigma_{0,1}(\Gamma')] = \sigma_{0,1}(z')) \; .
\end{equation}
For all $z'' \in \calS[\Gamma'']$,
\begin{equation}
\label{eq:vstationary}
\Prob(Y[\Gamma''] = z'') = \Prob(Y[\sigma_{1,-t}(\Gamma'')] = \sigma_{1,-t}(z'')) \; .
\end{equation}
\vspace{1ex}%
\makebox[0ex]{}\hrulefill\makebox[0ex]{}
\bbcaption{התוכנית הלינארית}{Linear program. The minimum (maximum) value is denoted $\lpmin^{*}$ ($\lpmax^{*}$) and is a lower (upper) bound on $R(\encoder)$.}{Linear program.}
\label{fig:linearProgram}
\end{figure}

In order to prove the theorem, we first state and prove a lemma, on a slightly modified linear program.

\begin{lemm}
\label{lemm:linearOne}
Fix $k > 0$, and replace (\ref{eq:hstationary}) and (\ref{eq:vstationary}) in Figure~\ref{fig:linearProgram} by
\[
\Big|\Prob(Y[\Gamma'] = z') - \Prob(Y[\sigma_{0,1}(\Gamma')] = \sigma_{0,1}(z')) \Big| \leq \frac{1}{k}
\]
and
\begin{multline*}
\Big|\Prob(Y[\Gamma''] = z'') - \Prob(Y[\sigma_{1,-t}(\Gamma'')] = \sigma_{1,-t}(z''))\Big|
\\
\leq \frac{t+1}{k} \; ,
\end{multline*}
respectively.

Denote the minimum and maximum of the resulting linear program as $\lpmin^{(k)}$ and $\lpmax^{(k)}$, respectively. Then,
 \[
\lpmin^{(k)} \leq R(\encoder) \leq \lpmax^{(k)} \; .
\]
\end{lemm}


\begin{proof}
Consider $\encoder^{(k)}$ (as defined by (\ref{eq:Ek})). For given $M$ and $N$, define the index sets
\[
\wasNabla = \partial(\Bis_{M,N}, \Lambda) \; , \quad \wasBarNabla = \bar\partial(\Bis_{M,N}, \Lambda) \; .
\]
Obviously,
\begin{equation}
\label{eq:Ubig}
\lim_{M,N \to \infty} \frac{\size{\wasBarNabla}}{M \cdot N} = 1 \; .
\end{equation}
Denote $A^{(k)} = A^{(k)}(\encoder, M,N)$. By (\ref{eq:Ubig}) and Lemma~\ref{lemm:AkversusAEk},
\[
R(\encoder^{(k)}) = \lim_{M,N \to \infty} \frac{H(A^{(k)}[\wasBarNabla]|A^{(k)}[\wasNabla])}{\size{\wasBarNabla}} \; .
\]
Notice that $\Psi \subseteq \Lambda$. Thus, $\wasBarNabla \subseteq \bar\partial_{M,N}$, and we have
\begin{align*}
H(A^{(k)}[\wasBarNabla]|A^{(k)}[\wasNabla]) = &\sum_{(i,j) \in \wasBarNabla} H(A_{i,j}^{(k)} | A^{(k)}[\Tis_{i,j} \cap \Bis_{M,N}]) \\
= & \sum_{(i,j) \in \wasBarNabla} H(A_{i,j}^{(k)} | \tau_{i,j}(A^{(k)},\Psi)) \; ,
\end{align*}
where $\Tis_{i,j}$ is as defined in (\ref{eq:Tij}) and the last equality follows from (\ref{eq:Aprob}).

We now prove the following claim: for all $(i,j) \in \wasBarNabla$, we have that
\begin{equation}
\label{eq:lpminkLowerBound}
\lpmin^{(k)} \leq H(A_{i,j}^{(k)} | \tau_{i,j}(A^{(k)},\Psi)) \; .
\end{equation}
To see this, fix some $(i,j) \in \wasBarNabla$, and define for all $z \in \calS[\Gamma]$,
\[
p^{(k)}(z) = \Prob(\tau_{i,j}(A^{(k)},\Gamma) = z ) \; .
\]
Substituting $\pi(z) = p^{(k)}(z)$, the objective function in Figure~\ref{fig:linearProgram} is equal to $H(A_{i,j}^{(k)} | \tau_{i,j}(A^{(k)},\Psi))$. Also, notice that the probability distribution of $Y$ is equal to that of $\tau_{i,j}(A^{(k)}, \Lambda)$. By the fact that $A^{(k)}$ is quasi-stationary (and thus, so is every sub-configuration of it), all the linear requirements in the modified linear program are satisfied (i.e., the $p^{(k)}(z)$'s form a feasible solution). So, our claim (\ref{eq:lpminkLowerBound}) is proved.

We conclude that $\lpmin^{(k)} \leq R(\encoder^{(k)})$. Thus, by Lemma~\ref{lemm:REequalToREk},
\[
\lpmin^{(k)} \leq  R(\encoder) \; .
\]
A similar proof yields $R(\encoder) \leq \lpmax^{(k)}$.
\end{proof}


\begin{proof}[Proof of Theorem~\ref{theo:upperLowerBound}]
First, note that the modified linear program defined in Lemma~\ref{lemm:linearOne} has at least one feasible solution, $p^{(k)}(z)$, whenever
$M$ and $N$ are large enough so that $\wasBarNabla$ is non-empty.

For a given $k$, denote the minimizing variable values of the modified linear program by $\pi^{(k)}(z)$, $z \in \calS[\Gamma]$. Think of these variable values as a vector
\[
\bfpi^{(k)} = (\pi^{(k)}(z))_{z \in \calS[\Gamma]} \; .
\]
By compactness, the series $\bfpi^{(k)}$, $k = 1,2,\ldots$, has a cluster point, which we denote by $\bfpi^*$. Obviously, $\bfpi^*$ implies a feasible solution for the linear program in Figure~\ref{fig:linearProgram}. More so, we must also have that the value of the objective function for this feasible solution is a lower bound on $R(\encoder)$. So,
\[
\lpmin^{*} \leq R(\encoder) \; .
\]
Similarly, we deduce that
\[
R(\encoder) \leq \lpmax^{*} \; .
\]
\end{proof}

\emph{Remark:} While the definition of the encoder $\encoder$ includes (besides $\Psi$ and $\mu$) also the boundary distributions $\bfdelta = (\delta_{M,N})_{M,N>0}$, the bounds $\lpmin^{*}$ and $\lpmax^{*}$ do \emph{not} depend on $\bfdelta$.

Applying Theorem~\ref{theo:upperLowerBound} to our running example, with $r=4$, $s=5$, $t=1$, gives
\[
0.42430953 \leq R(\encoder) \leq 0.42442765 \; .
\]
To the best of our knowledge, our running example is the highest rate bit-stuffing encoder known, given that we are allowed to use at most two coins (i.e., two probability transformers). For comparison, we have calculated by the method presented in \cite{CalkinWilf:97} that
\[
\capacity(\calS_\sq) \leq 0.425078 \; .
\]
Namely, with two coins we achieve a rate that is only $0.2\%$ less than capacity.


Table~\ref{tbl:oneCoin} contains our results for a number of constraints. We abbreviate the ``no isolated bits'' constraints as ``n.i.b.''. In the first three rows, we compare ourselves to the results in \cite{Halevy+:04} (Table 1 and Equation (12)). For the comparison to be fair, we restrict ourselves to the neighbor sets $\Psi$ used in \cite{Halevy+:04}, and use the same number of coins.

\begin{table}
\small
\caption{Bounds on the rates of encoders using 
a small number of coins.}
\begin{center}
\begin{tabular}{||c|c|c|c|c||}
\hline
Constraint & Coins & $\lpmin^*$ & $\lpmax^*$ & \cite{Halevy+:04} \\
\hline
$(2, \infty)$-RLL &  1 &   0.440722  & 0.444679 & 0.4267 \\
\hline
$(3,\infty)$-RLL & 1 &  0.349086 & 0.386584 & 0.3402 \\
\hline
n.i.b. & 2 &   0.917730 &  0.919395 & 0.9127 \\
\hline \hline
$(1, \infty)$-RLL &  3 &   0.587776  & 0.587785 & --- \\
\hline
\end{tabular}
\end{center}
\label{tbl:oneCoin}
\vspace{-15pt}
\end{table}

\paperonly{\section{A lower bound on capacity}}%
\thesisonly{\bsection{חסם תחתון על קיבול}{A lower bound on capacity}}%
\label{sec:lowerBoundOnCapacity}
The following is a straightforward corollary of Theorem~\ref{theo:upperLowerBound}.
\begin{coro} For every bit-stuffing encoder $\encoder$,
\[
\lpmin^*(\encoder) \leq \capacity(\calS) \; .
\]
\end{coro}
Thus, we can use the minimizing linear program of Figure~\ref{fig:linearProgram} to bound $\capacity(\calS)$ from below.

To obtain better lower bounds on $\capacity(\calS)$, we can search for good $\Psi$ and $\mu$. For instance, for the set $\Psi=\Psi_\sq$ in (\ref{eq:PsiRunning}), the function $\mu_\sq$ in (\ref{eq:musq}) was obtained by maximizing $\lpmin^*$ over all $\mu$ that form with $\Psi_\sq$ (and every $\bfdelta$) a bit-stuffing encoder for $\calS_\sq$. Better lower bounds can be obtained by looking at larger sets $\Psi$ (at the price of higher computational complexity).

Table~\ref{tbl:manyCoins} summarizes our results for certain constraints.
The last two columns contain previously published lower bounds on the capacity of the corresponding constraint. We have highlighted values of $\lpmin^*$  which are an improvement of these previously known results. The bounds in the penultimate column are taken from \cite{SharovRoth:08}, which was published recently. We note that the method used in \cite{SharovRoth:08} is quite different than ours. As can be seen, both \cite{SharovRoth:08} and our method are comparable. The bounds in the last column are taken from \cite{OrdentlichRoth:07}, \cite{ForchhammerLaursen:07}, \cite{AshleyMarcus:98}, and \cite{ForchhammerLaursen:07a}, respectively: they were the the best known when our method was first published in \cite{TalRoth:08} (at the same time as \cite{SharovRoth:08}).

\begin{table}
\small
\caption{Bounds on the rates of certain bit-stuffing encoders.}
\begin{center}
\begin{tabular}{||c|c|c|c|c|c||}
\hline
Constraint & Coins & $\lpmin^*$ & $\lpmax^*$ & \cite{SharovRoth:08} & Others \\
\hline
$(2,\infty)$-RLL & 5 & \textbf{0.44420} & 0.4450 & 0.44417 & 0.4423 \\
\hline
$(3, \infty)$-RLL &  2 &     0.35973 & 0.3690 & 0.36562 & 0.3641 \\
\hline
$(0,2)$-RLL & 66 &     0.81549 & 0.8169 & 0.81600 & 0.7736 \\
            & 18 &     0.81501 & 0.8162 &  &\\
            & 9  &     0.81073 & 0.8197  & & \\
\hline
n.i.b. & 56 & \textbf{0.92264} & 0.9238 & 0.92086 & 0.9156 \\
\hline
\end{tabular}
\end{center}
\label{tbl:manyCoins}
\vspace{-15pt}
\end{table}
%
%
\section*{Acknowledgment}
The first author wishes to thank Roee Engelberg for very stimulating discussions.
\twobibs{
\bibliographystyle{IEEEtran}
\bibliography{../../mybib}
}
{

}
\end{document}